# Theoretical Study of Charge Transport Properties of Curved PAH Organic Semiconductors

Heng-yu Jin[a†], Xiao-qi Sun[a,c†], Gui-ya Qin[b], Zhi-peng Tong[a], Rui Wang[b], Qi Zhao[a], Ai-Min Ren[b], Jing-fu Guo*[a]

Curved polycyclic aromatic hydrocarbons (PAHs) exhibit distinctive geometric and electronic structures, rendering them highly promising in addressing issues of solubility and air stability, which are faced for large linear arene π-conjugated organic semiconductors. In this study, a series of surface-curved PAHs and the heteroatom doped derivatives are selected and designed, and the relationship between electronic structure and charge transport properties of these molecules is investigated by using density functional theory (DFT). And the effects of sulfur/oxygen, nitrogen and boron doping on the charge transport performance of curved PAH semiconductors are explored. The results show that curved PAHs exhibit improved solubility and stability, with the degree of molecular curvature significantly affecting the material's transport properties. Six-membered ring PAHs with deeper bowl-like structures (d > 1.0 Å) and their N/B dopants tend to form quasi-one-dimensional, slightly sliding, compact π-stack structures with concave-to-convex configurations, exhibiting superior hole transport properties compared to those with shallower bowl-like structures (0.5 Å < d <1.0 Å) and loose stacks (**B2**, **B4**, **B6**). S/O doping between benzene rings to form seven-membered ring can significantly reduce the bowl shaped depth (d < 1.0 Å), but increases reorganization energy, making the PAH easier for 2D π-π stacking; However, further doping N/B atom at the edges or core of the PAHs can fine-tune the bowl shaped depth, and suppress the increase in hole reorganization energy caused by S/O doping, maintaining the hole reorganization energy of heteroatom-doped PAHs ca. 200 meV, which is obliged for high mobility materials. Introducing S/O/N atoms can increase the bandgap and enhance the optical stability of PAHs. Incorporating sulfur atom (which inhibit intermolecular rotation stacking) and boron atom (which increase intermolecular overlapping integral and transfer integral), **B5** achieves a significant increase in hole mobility (3.49 $cm^2V^{-1}s^{-1}$).

.

## Introduction

Organic semiconductor materials (OSCs) exhibit remarkable application potential across a wide range of fields, including sensing technologies, energy conversion, information display, wearable devices, and electronic skins. This is attributed to their unique advantages, such as high flexibility, ultra-thinness, low energy consumption, and large-area processability.[1-4] Optoelectronic devices based on OSCs, such as organic field-effect transistors (OFETs), organic photovoltaics (OPVs), organic light-emitting diodes (OLEDs), and organic light-emitting transistors (OLETs), have attracted extensive attention due to their wide-ranging applicability.[5-7] However, further enhancement of device performance fundamentally relies on improving the charge transport efficiency of organic semiconductors, particularly through the rational design and synthesis of materials with high carrier mobility—while the underlying theoretical frameworks for such improvements remain insufficiently developed. For OFETs, higher carrier mobility directly translates to shorter switching times. This attribute is of paramount importance in the communications industry, where it can accelerate data transmission rates to enable faster and more stable data communication. In computing technology, enhanced mobility facilitates quicker processor operations, allowing computer systems to execute diverse tasks with greater efficiency. Additionally, high-mobility OFETs play a critical role in application domains such as electronic display manufacturing, automotive electronics development, and aerospace technology. Currently, most π-conjugated organic molecules employed as P-type semiconductor materials have garnered significant attention due to their superior charge transport properties. Nevertheless, persistent challenges in solubility and stability require urgent resolution.[8] Overcoming this bottleneck will significantly enhance the performance of optoelectronic devices, laying a solid foundation for the widespread practical application of organic semiconductor materials. Large-scale organic π-conjugated molecules are pivotal to high-performance optoelectronic materials, as their strong intermolecular π-conjugation coupling provides a foundation for efficient charge carrier transport. For instance, among linear acenes such as anthracene, tetracene, and pentacene, pentacene-based OFETs exhibit a remarkable hole mobility of up to 40 $cm^2V^{-1}s^{-1}$.[9] However, extending the π-conjugation system in straight chains by increasing the number of fused aromatic rings leads to significant decreases in solubility and stability, which has become a critical bottleneck for their development. PAHs and their derivatives have attracted considerable attention due to their exceptional optoelectronic properties in organic electronic devices.[10-13] For instance, pyrene[14] and perylene[15] used in OFETs demonstrate promising applicable prospects. Compared to linear acenes, PAHs possess larger π-

[a.] School of Physics, Northeast Normal University, Changchun, 130024, China.
[b.] Institute of Theoretical Chemistry, College of Chemistry, Jilin University, Changchun, 130023, China.
[c.] Department of Physics, College of Science, Yanbian University, Yanji, 133002, China.
* Corresponding author：Jing-Fu Guo, E-mail address: guojf217@nenu.edu.cn
† Heng-yu Jin and Xiao-qi Sun contributed equally to this work.

Supplementary Information available: [details of any supplementary information available should be included here]. See DOI: 10.1039/x0xx00000x

conjugated systems, which may generate stronger electronic coupling effects that facilitate charge transport, making them ideal candidates for organic semiconductor devices.[16] Additionally, PAHs avoid the diminished photochemical and air stability issues that arise with the increase in aromatic rings in linear acenes.[17] Nonetheless, challenges persist, such as strong intermolecular repulsion that impedes the formation of efficient charge transport channels and poor solubility in higher-molecular-weight PAHs. PAHs offer notable advantages in addressing the poor solubility and stability issues inherent in planar PAH-based organic semiconductors, primarily due to their unique structural and optoelectronic characteristics.[18,19] As the first synthesized bowl-shaped π-conjugated molecule, corannulene was developed by Barth and Lawton[20] (FIGURE 1a) and has emerged as a pivotal material in the field of organic light-emitting diode (OLED) devices.[21] The study of sumanene represents the second major breakthrough in the domain of bowl-shaped π-conjugated molecules (FIGURE 1a).[22,23] Seki and Hiro first demonstrated that sumanene exhibits anisotropic charge transport properties: its ordered columnar π-π stacking structure achieves a hole mobility of 0.75 $cm^2V^{-1}s^{-1}$,[24] thereby establishing a foundation for subsequent applications. However, conventional corannulene has limited utility as OSCs due to the absence of π-π stacking interactions. Studies have demonstrated that introducing functional groups at the molecular periphery can modulate its electronic structure, inducing a concave-convex one-dimensional columnar stacking mode[25]. This structural adjustment significantly enhances charge carrier transport performance.[25-28] Currently, a series of bowl-shaped polycyclic aromatic hydrocarbons (PAHs) have exhibited promising potential in organic semiconductor applications.[19,29-37] For example, triselenasumanene exhibits a mobility of 0.37 $cm^2V^{-1}s^{-1}$ in organic transistors, while functionalized 1,2-bis(CF$_3$)-Corannulene achieves an electron mobility of 0.9 $cm^2V^{-1}s^{-1}$. However, certain curved molecules often suffer from poor charge transport due to challenges in forming tight π-π stacking structures, which remains a critical technical bottleneck demanding urgent breakthroughs in this field.

Heteroatom functionalization has unlocked new avenues for curved PAHs to serve as high-mobility materials. Experimental investigations have shown that incorporating heteroatoms into PAHs can significantly modulate their electronic structures and physicochemical properties, thereby enhancing both air stability and charge transport performance.[17] The introduction of heteroatoms into π-conjugated frameworks imparts materials with distinctive optoelectronic properties.[19,38] Particularly in bowl-shaped PAH systems, the modulation effects of heteroatoms on electronic structures and their profound influence on molecular properties have garnered substantial attention. Notably, synthetic research on heteroatom-doped derivatives of Corannulene and Sumanene has achieved remarkable advancements.[39]

The research group led by Shingo Ito has achieved notable progress in the synthesis and optoelectronic characterization of curved PAHs and their heteroatom-doped derivatives. In 2015, the group successfully synthesized a deeply bowl-shaped PAH molecule (**A1** in FIGURE 1b), characterized by a unique one-dimensional concave-convex π-π stacking structure. This achievement established a foundation for subsequent research in this field.[40] Furthermore, in 2022, the group designed and synthesized a series of seven-membered heterocyclic PAHs (**B1** and **B2** in FIGURE 1b). Among them, the sulfur-containing seven-membered heterocyclic PAHs exhibit a quasi-planar conformation, featuring a unique pitched π-π stacking mode in their crystalline state. This gives rise to significant C-H···π interactions and π-π orbital overlap between molecules. Single-crystal field-effect transistors fabricated from these molecules exhibit exceptional P-type charge transport properties, with hole mobility reaching up to 1.05 $cm^2V^{-1}s^{-1}$—two orders of magnitude higher than that of THA1 (FIGURE 1a).[16] In 2023, the group reported the synthesis of pyridine-fused azabowl PAHs (**A2** and **A3** in FIGURE 1b) and investigated the effects of nitrogen atoms introduced at the periphery on their structural, electronic, and optical properties.[38] Current research on the charge transport properties of curved bowl-shaped PAHs remains incomplete, particularly regarding the lack of fundamental principles for designing high-mobility curved PAH-based organic semiconductors. In this study, a series of curved PAH derivatives (**A1–A4** and **B1–B6**, as shown in FIGURE 1b,c) were selected and designed following a naming convention: six-membered ring-containing PAHs are categorized as the A-series (**A1–A4**) (FIGURE 1b,c), and seven-membered heteroring-containing PAHs as the B-series (**B1–B6**) (FIGURE 1b,c). In this study, the relationships between the electronic structures of curved PAHs and charge transport performance will be investigated using density functional theory (DFT). The focus will be on elucidating the effects of sulfur/oxygen, nitrogen, or boron doping on critical charge transport parameters, such as molecular stacking, transfer integrals, and reorganization energies.

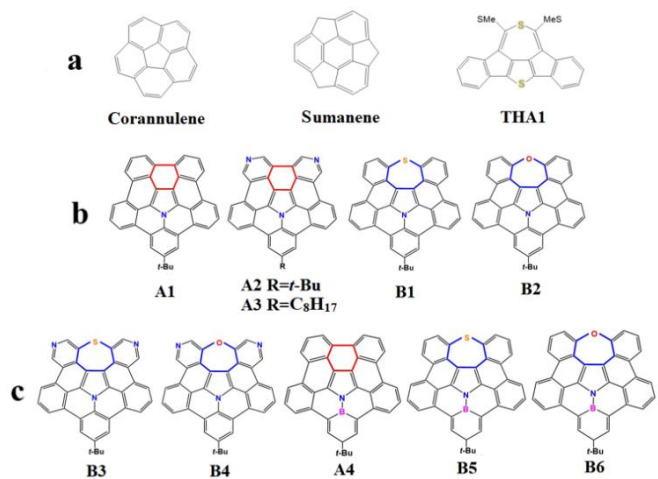

FIGURE 1. a. Corannulene, Sumanene, THA1, b. **A1**, **A2**, **A3**, **B1** and **B2**, c. **A4**, **B3**, **B4**, **B5** and **B6**.

This study mainly focuses on the following three aspects.

(1) Influence of heteroatom doping on molecular curvature and molecular stacking configurations.

(2) Impact of heteroatom doping on frontier molecular orbitals, ionization energies, electron affinity potentials, and charge injection barriers.

(3) Effect of heteroatom doping on frontier molecular orbital distributions, reorganization energies, and charge transfer integrals.

## Theoretical calculation methods

### Charge carrier transport mechanism and calculation method

Organic semiconductor molecules are primarily bound by van der Waals forces, leading to larger intermolecular distances, weaker intermolecular interactions, and reduced electronic coupling. Leveraging this characteristic, the charge transfer process between lattice sites can be described using a hopping model[41], with its charge transfer rate quantified by the Marcus formula (see Equation (1)).[42]

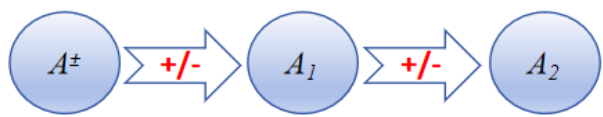

FIGURE 2. Jump mechanism.

$$k_{CT} = \frac{|V|^2}{\hbar}\sqrt{\frac{\pi}{\lambda k_B T}} exp\left(-\frac{\Delta G_0+\lambda}{4k_B T}\right) \quad (1)$$

According to Marcus theory, the charge transfer rate is primarily governed by the internal reorganization energy ($\lambda$) and the transfer integral ($V$). Internal reorganization energy refers to the geometric configuration relaxation energy of a molecule during its transition from a neutral state to an ionic state; smaller values are more favorable for enhancing charge transfer efficiency. Typically, the internal reorganization energy is quantitatively evaluated using two methods: (1) the adiabatic potential energy surface method (Equation (2))[43,44] and (2) the normal mode analysis method (Equation (3)). The latter is only applicable to harmonic oscillator systems.

$$\lambda_{h/e} = E_{h/e}(Q_N) - E_{h/e}(Q_{h/e}) + E_N(Q_{h/e}) - E_N(Q_N) \quad (2)$$

$$\lambda = \sum_j \lambda_j \sum_j S_j \hbar\omega_j \quad (3)$$

The transfer integral ($V$), a critical parameter for assessing the strength of electronic coupling during charge transfer, is typically evaluated using the lattice energy correction method (Equation (4)), with results from this method widely acknowledged.[45-47] In this study, the transfer integral values were calculated using the AMS software package at the PW91/TZP theoretical level[48], a combination proven capable of reasonably assessing the electronic coupling strength between organic crystal molecules.[49]

$$V_{if} = \frac{H_{if} - \frac{1}{2}S_{if}(H_{ii}+H_{ff})}{1-S_{if}^2} \quad (4)$$

The carrier mobility ($\mu$), which quantifies the average drift velocity of electrons per unit electric field strength, serves as a critical performance metric for organic field-effect transistor materials. As described by Einstein's equation (see Equation (5))[43], the carrier mobility can be calculated as follows:

$$\mu = \frac{eD}{k_B T} \quad (5)$$

Among these, the diffusion coefficient $D$ is determined via Monte Carlo simulation (see Equation (6)). To ensure the convergence of the diffusion coefficient, 2000 Monte Carlo simulations were performed for each transport channel to derive a linear relationship between the mean square displacement (MSD, $\langle x^2(t)\rangle$) and diffusion time ($t$). The diffusion coefficient $D$ is ultimately calculated by linearly fitting this relationship.[44]

$$D = \frac{1}{2n}\lim_{t\to\infty}\frac{\langle x(t)\rangle^2}{t} \quad (6)$$

**Calculation level and software**

The initial molecular structures utilized in this study were derived from X-ray diffraction analysis data obtained from the Cambridge Crystallographic Data Centre (CCDC). To characterize the system, the following methods were employed.

First, the hydrogen atom positions were optimized using the VASP package[50,51], employing the PBE functional[52] and PAW pseudopotential method.[53] Subsequently, the geometric structures of neutral and ionic states were optimized at the B3LYP/6-311G(2d,p) level using Gaussian09 software[54], with vibrational frequency analysis confirming that the structures corresponded to potential energy surface minima.

Next, solvation free energy ($\Delta G_{sol}$) was calculated based on the SMD model.[55] Meanwhile, the internal reorganization energy was determined using the adiabatic potential energy surface (APES) method and the normal mode analysis (NMA) approach from the CTMP package.[56]

Furthermore, the transfer integral was obtained via the lattice site energy correction method. These results were then integrated with Marcus theory and kinetic Monte Carlo simulations to compute charge transfer rates and diffusion coefficients, with mobility subsequently derived using the Einstein relation.

Additionally, Hirshfeld surface analysis was performed using Crystal Explorer[57], and intermolecular interactions were investigated via the SCS-SAPT0/jun-cc-pVDZ method implemented in the PSI4 package.[58]

## Results and discussion

### Geometric structure, solubility, and electronic structure of molecules

#### Bowl shaped depth and key angles of molecules

Based on the single-crystal structure, the ground-state structures of the molecules were optimized. Considering that the geometric structure of a molecule may be influenced by its surrounding stacking environment—particularly the spatial constraints of nonplanar molecules—the quantum mechanics/molecular mechanics (QM/MM) model was employed to simulate geometric structure optimization in the solid phase. Results indicated that structures simulated via QM/MM more accurately reflected crystal structure characteristics than those in the gas phase. Key angles in the optimized molecular structures are listed in TABLE S1 (Supplementary Information), with atomic numbering illustrated in FIGURE 3.

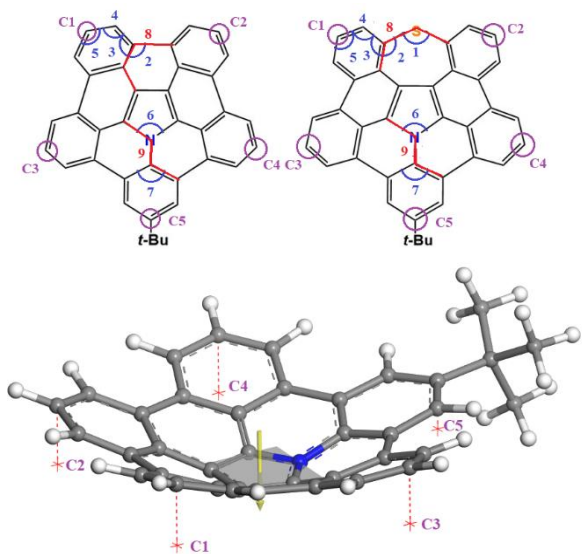

FIGURE 3. Schematic diagram of partial atomic numbering, key angles, and bowl shaped depth of the studied molecule.

First, the effect of heteroatom introduction on the bowl depth of the molecules was analyzed. The bowl depth is defined as the average distance between the five carbon atoms (C1, C2, C3, C4, and C5) and the plane of the pyrrole ring, as illustrated in FIGURE 3, with results tabulated in TABLE 1.

**TABLE 1** The bowl shaped depth of the studied molecules (unit: Å).

| Molecules | Distances from each vertex of the molecule to the plane of the pyrrole ring | Bowl shaped depth |
|---|---|---|
| A1 | 1.296-1.657 | 1.523 |
| A2 | 1.212-1.748 | 1.550 |
| A3 | 1.443-1.709 | 1.566 |
| A4 | 1.398-1.762 | 1.575 |
| B1 | 0.043-0.472 | 0.228 |
| B2 | 0.772-0.868 | 0.831 |
| B3 | 0.024-0.283 | 0.137 |
| B4 | 0.821-0.934 | 0.867 |
| B5 | 0.049-0.472 | 0.276 |
| B6 | 0.441-0.839 | 0.713 |

Based on TABLE 1, the studied molecules can be categorized into three groups according to their bowl depth: A-series molecules exhibit deeper bowl structures (bowl depth > 1.5 Å), B-series molecules feature quasi-planar structures (bowl depth < 0.5 Å), and **B2**, **B4**, and **B6** display shallower bowl structures (0.5 Å < bowl depth < 1.5 Å). TABLE 1 reveals that heteroatom introduction significantly influences bowl depth: specifically, compared to six-membered ring A-series molecules (1.523–1.575 Å), seven-membered ring B-series molecules (0.137–0.867 Å) prepared by introducing sulfur or oxygen atoms between rings show reduced bowl depth. This is primarily attributed to the formation of seven-membered rings via the introduction of sulfur or oxygen atoms between rings, which exhibit lower ring strain; the dihedral angles of both carbon-sulfur and carbon-oxygen bonds in B-series molecules are relatively large, confirming the presence of low ring strain. Except for **B1**, the symmetrical introduction of nitrogen atoms into the peripheral benzene ring significantly increases the bowl depth of the molecule. Oxygen-containing molecules exhibit greater bowl depth than sulfur-containing analogs, which is closely related to the smaller atomic radius of oxygen, inducing larger ring strain. For example, compared to **A1**, the introduction of nitrogen atoms into the peripheral benzene ring to form **A2** and **A3** increases Angle 4 (123.9°/123.8° vs. 120.5°) while decreasing Angle 5 (120.3°/120.4° vs. 122.7°). This effect arises from the larger atomic radius of nitrogen compared to carbon, leading to increased tension within the six-membered ring. Consequently, the bowl depth increases significantly upon nitrogen substitution. In contrast, introducing boron atoms into the center of **A1**, **B1**, and **B2** to form **A4**, **B5**, and **B6** increases Angle 7, resulting in deeper bowls. This is attributed to the smaller radius of boron compared to carbon, forming shorter boron-nitrogen bonds than carbon-nitrogen bonds and thereby increasing tension within the five-membered ring. At the boron substitution site, the dihedral angle (Angle 9) formed by the boron-nitrogen bond with adjacent carbon atoms remains nearly unchanged, with sulfur-containing molecules exhibiting larger Angle 9 values than oxygen-containing counterparts.

**Solubility**

The solubility of organic charge-transport materials is a critical physical property that directly influences the processability and applicability of organic semiconductor materials in electronic device fabrication. Thus, molecules designed for charge-transport applications must exhibit not only excellent intrinsic properties but also favorable solubility for practical implementation. The solvation free energy ($\Delta G_{sol}$) serves as a metric to quantify the solubility of solutes in solvents. The more negative the solvation free energy value, the higher the solubility of molecules in the solvent.[59] TABLE 2 lists the solvation free energies ($\Delta G_{sol}$) of the studied molecules in water, dichloromethane, and chloroform—calculated using the SMD solvent model at the M052X/6-31G(d) level—and compares them with those of pentacene, a planar linear structure, in these solvents.

**TABLE 2** At the M052X/6-31G (d) level, the solvent model uses the solvation free energy ($\Delta G_{sol}$) (unit: kcal/mol) of pentacene and the studied molecules in common solvents: water, dichloromethane and chloroform calculated by SMD.

| Molecules | Water | Dichloromethane | Chloroform |
|---|---|---|---|
| Pentacene | -6.87 | -12.84 | -18.13 |
| A1 | -8.76 | -27.83 | -25.86 |
| A2 | -13.78 | -27.93 | -25.79 |
| A3 | -12.64 | -30.68 | -28.68 |
| A4 | -9.02 | -27.11 | -25.02 |
| B1 | -9.10 | -28.67 | -26.64 |
| B2 | -8.75 | -27.42 | -25.42 |
| B3 | -13.70 | -28.29 | -26.17 |
| B4 | -13.41 | -27.25 | -25.10 |
| B5 | -9.19 | -27.87 | -25.73 |
| B6 | -8.70 | -26.48 | -24.41 |

Systematic analysis of solubility revealed that the studied molecules generally exhibit better solubility than the planar linear structure of pentacene. Notably, all molecules show higher solubility in organic solvents than in water.

When comparing A-series and B-series molecules, their solubility in water is comparable; in organic solvents, B-series molecules with sulfur atoms exhibit better solubility than A-series, whereas B-series molecules with oxygen atoms show weaker solubility than A-series. These findings demonstrate that the designed molecules possess excellent solubility in both aqueous and organic solvent systems, providing a critical foundation for their practical applications.

**Ionization potential, electron affinity, frontier molecular orbital HOMO, LUMO, and HOMO-LUMO bandgap energy**

Charge carrier injection significantly influences the performance of field-effect transistor devices. The efficiency of charge injection is governed by frontier molecular orbitals, ionization potential (IP), and electron affinity (EA).

The calculated values of IP, EA, highest occupied molecular orbital (HOMO) energy level, lowest unoccupied molecular orbital (LUMO) energy level, and energy gap ($E_{gap}$) are presented in TABLE 3. The distributions of HOMO and LUMO for the studied molecules are illustrated in FIGURE 4 and FIGURE 5.

TABLE 3 The ionization potentials (IPs), electron affinities (EAs), HOMO levels, LUMO levels, and energy gap values ($E_{gap}$) of the studied molecules calculated at the B3LYP/6-311G(2d,p) level (unit: eV).

| Molecules | AIP | AEA | VIP | VEA | HOMO | LUMO | $E_{gap}$ |
|---|---|---|---|---|---|---|---|
| **A1** | 6.11 | 0.65 | 6.17 | 0.56 | -5.00 | -1.73 | 3.27 |
| **A2** | 6.50 | 0.93 | 6.57 | 0.85 | -5.38 | -2.02 | 3.37 |
| **A3** | 6.50 | 0.93 | 6.57 | 0.85 | -5.38 | -2.02 | 3.36 |
| **A4** | 6.04 | 1.08 | 6.11 | 0.94 | -4.94 | -2.08 | 2.87 |
| **B1** | 6.28 | 0.81 | 6.37 | 0.72 | -5.21 | -1.86 | 3.36 |
| **B2** | 6.30 | 0.73 | 6.39 | 0.65 | -5.23 | -1.80 | 3.43 |
| **B3** | 6.66 | 1.11 | 6.76 | 1.01 | -5.57 | -2.19 | 3.38 |
| **B4** | 6.72 | 1.02 | 6.80 | 0.94 | -5.63 | -2.09 | 3.53 |
| **B5** | 6.36 | 1.12 | 6.44 | 1.03 | -5.27 | -2.16 | 3.11 |
| **B6** | 6.27 | 1.04 | 6.35 | 0.93 | -5.19 | -2.07 | 3.12 |

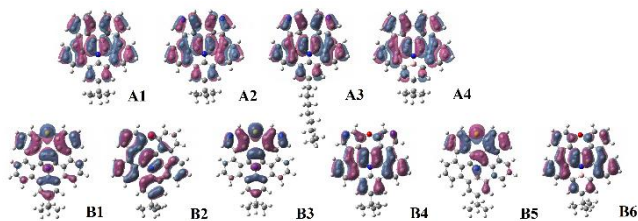

FIGURE 4. The HOMO electron density distribution of the studied molecules.

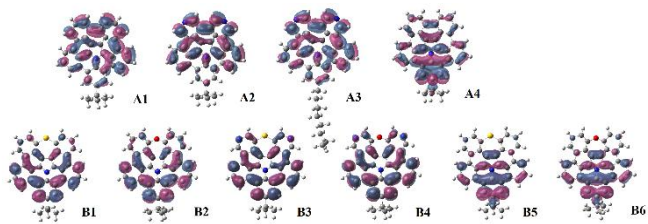

FIGURE 5. The LUMO electron density distribution of the studied molecules.

IP and EA are critical parameters that respectively characterize the reducing and oxidizing capabilities of molecules. The theoretical predictions of IP and EA were obtained without accounting for interfacial effects and the polarization environment at the organic–metal interface, leading to a tendency for theoretical IP values to be overestimated and EA values to be underestimated. Despite potential discrepancies between these theoretical values and experimental reality, they still reflect the qualitative trends in material property changes to a significant extent.

The higher the IP of a material molecule, the more resistant it is to oxidation and the stronger its oxidation stability.[60] Compared with **A1**, the IP values of **A2** and **A3** increase; analogously, the IP values of **B3** and **B4** are higher than those of **B1** and **B2**, respectively. These results suggest that the introduction of nitrogen atoms enhances the oxidation stability of the molecules.

The energy difference between the lowest unoccupied molecular orbital LUMO and HOMO, termed the $E_{gap}$, can reflect a material's optical stability to a certain extent. A larger $E_{gap}$ corresponds to better optical stability. Calculations at the B3LYP/6-311G(2d,p) level show that the HOMO energy levels of A-series and B-series molecules range from −5.63 to −4.94 eV, while their LUMO energy levels span −2.19 to −1.73 eV, both demonstrating stable hole-transport properties.

After the introduction of nitrogen atoms, the $E_{gap}$ of both A-series and B-series molecules increase, suggesting that nitrogen incorporation enhances the optical stability of the molecules. Conversely, introducing boron atoms to form **A4**, **B5**, and **B6** reduces the energy gap. The IP and $E_{gap}$ of seven-membered ring B-series molecules are greater than those of their corresponding six-membered ring A-series counterparts, with oxygen-containing seven-membered ring molecules exhibiting larger $E_{gap}$ values than sulfur-containing analogs. These results indicate that B-series molecules with heteroatom-incorporated seven-membered rings between rings possess stronger oxidation and optical stability compared to A-series molecules with intra-ring doping. Among them, oxygen-containing seven-membered ring molecules show relatively stronger optical stability than sulfur-containing ones.

The HOMO and LUMO electron density distributions of the studied molecules are shown in FIGURE 4 and FIGURE 5. The molecules exhibit symmetry differences between HOMO and LUMO: HOMO is distributed along the molecular principal axis ($C_2$ symmetry axis), whereas LUMO is oriented perpendicular to this axis. For A-series molecules, both HOMO and LUMO show delocalized and complementary distributions, while in B-series molecules, LUMO localizes in the lower half of the pyrrole ring and HOMO predominantly distributes in the upper half. Specifically, A-series HOMOs are uniformly distributed within the polycyclic aromatic hydrocarbon framework (excluding the pyrrole axis); **B4** and **B6** with oxygen-containing seven-membered rings exhibit similar distributions to A-series but **B2** shows asymmetric polarization; and HOMOs of **B1**, **B3**, and **B5** with sulfur-containing seven-membered rings are concentrated on the upper part of the pyrrole ring and symmetrical rings. The incorporation of sulfur atoms contributes to HOMO construction, whereas oxygen atoms do not participate in HOMO formation. The distinct frontier molecular orbital characteristics induced by different dopant atoms may give rise to variations in intermolecular stacking patterns and transfer integrals. Compared with **B4**, **B3** exhibits greater delocalization over the six-membered rings flanking the seven-membered ring. This phenomenon may arise from the higher electronegativity of nitrogen atoms introduced into the seven-membered rings of **B3** and **B4** compared to carbon, as well as the higher electronegativity of oxygen and sulfur atoms within the seven-membered ring, which collectively induce repulsive effects. These factors lead to more localized HOMO distributions on the seven-membered ring and adjacent six-membered rings in **B3** and **B4** compared to **B1** and **B2**. Moreover, due to the higher electronegativity of oxygen relative to sulfur, the HOMO localization degree in **B4** is greater than that in **B3**.

To perform quantitative analysis and uncover the underlying causes of HOMO distribution changes, we divided the molecular skeletons of A-series and B-series into seven primary fragments as depicted in FIGURE 6. Fragment 1 represents a pyrrole ring, Fragments 2–6 denote peripheral benzene rings, and Fragment 7 signifies a six-membered or seven-membered ring. Using Multiwfn software[61], the contribution of each molecular fragment to the HOMO was calculated via the Natural Atomic Orbital (NAO) method, with results presented in TABLE 4.

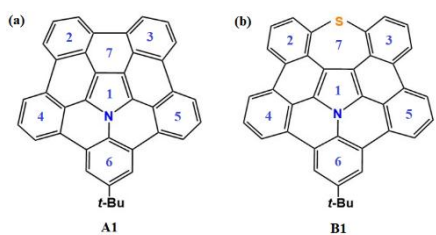

FIGURE 6. Schematic illustration of fragment division for the studied molecules (taking **A1** and **B1** as examples).

**TABLE 4** Contributions of molecular fragments to the HOMO (unit: %).

| Molecules | Frg1 | Frg2 | Frg3 | Frg4 | Frg5 | Frg6 | Frg7 |
|---|---|---|---|---|---|---|---|
| **A1** | 33.15 | 12.84 | 12.81 | 16.77 | 17.19 | 7.68 | 16.51 |
| **A2** | 32.96 | 12.11 | 11.98 | 17.28 | 17.19 | 8.47 | 18.14 |
| **A3** | 33.01 | 11.95 | 12.06 | 17.28 | 17.34 | 8.33 | 18.07 |
| **A4** | 34.82 | 12.63 | 13.12 | 16.65 | 17.08 | 5.55 | 17.34 |
| **B1** | 23.51 | 17.01 | 20.30 | 3.14 | 4.19 | 10.66 | 55.80 |
| **B2** | 32.14 | 23.51 | 3.98 | 15.94 | 7.65 | 13.51 | 30.80 |
| **B3** | 18.72 | 18.16 | 18.70 | 1.76 | 1.92 | 8.30 | 63.67 |
| **B4** | 35.56 | 7.60 | 5.50 | 20.28 | 19.25 | 11.72 | 18.95 |
| **B5** | 16.15 | 17.06 | 26.52 | 0.25 | 3.20 | 4.17 | 63.28 |
| **B6** | 38.34 | 7.21 | 8.03 | 19.12 | 17.44 | 7.22 | 18.80 |

Analysis of the HOMO distribution maps (FIGURE 4 and TABLE 4) shows that for A-series molecules, although the nitrogen atom in Fragment 1 (the pyrrole ring) contributes negligibly to the HOMO, the pyrrole ring exhibits the highest HOMO distribution proportion, which is nearly consistent across derivatives (**A1**: 33.15%; **A2**: 32.96%; **A3**: 33.01%; **A4**: 34.82%). Compared with **A1**, the introduction of nitrogen atoms in **A2** and **A3** leads to decreased HOMO distribution in Fragments 2 (**A1**: 12.84%; **A2**: 12.11%; **A3**: 11.95%) and 3 (**A1**: 12.81%; **A2**: 11.98%; **A3**: 12.06%), but increased distribution in Fragments 4 (**A1**: 16.77%; **A2**: 17.28%; **A3**: 17.28%), 5 (**A1**: 17.19%; **A2**: 17.19%; **A3**: 17.34%), 6 (**A1**: 7.68%; **A2**: 8.47%; **A3**: 8.33%), and 7 (**A1**: 16.51%; **A2**: 18.14%; **A3**: 18.07%). This indicates that the introduction of two nitrogen atoms enhances their distribution on the HOMO, while the nitrogen-incorporated rings (Fragments 2 and 3) show minimal HOMO contribution, with electron density redistributed toward the central hexagonal ring (Fragment 7). A comparison between **A1** and **A4** reveals that boron atom introduction reduces Fragment 6's HOMO contribution from 7.68% (**A1**) to 5.55% (**A4**).

For B-series molecules, the HOMO distribution on Fragment 1 (the pyrrole ring) in oxygen-containing seven-membered ring molecules is higher than that in sulfur-containing seven-membered ring analogs. Compared with **B1**, the introduction of nitrogen or boron atoms in **B3** and **B5** causes a decrease in HOMO distribution on Fragment 1 (**B1**: 23.51%; **B3**: 18.72%; **B5**: 16.15%) but an increase on Fragment 7 (**B1**: 16.51%; **B3**: 18.14%; **B5**: 18.07%). In contrast to **B1**, introducing nitrogen and boron atoms into **B4** or **B6** based on **B2** enhances the HOMO distribution on Fragment 1 (**B2**: 32.14%; **B4**: 35.56%; **B6**: 38.34%) while reducing it on Fragment 7 (**B2**: 30.80%; **B4**: 18.95%; **B6**: 18.80%). The introduction of nitrogen atoms also decreased the contributions of Fragments 2 and 3 to the HOMO (**B2**: 27.49%; **B4**: 13.10%; **B6**: 15.24%). A comparison of **B1**, **B2**, **B5**, and **B6** reveals that boron atom introduction reduces Fragment 6's HOMO contribution (**B1**: 10.66%; **B2**: 13.51%; **B5**: 4.17%; **B6**: 7.22%). The HOMO distribution over the sulfur-containing seven-membered ring (Fragment 7) and its adjacent rings (Fragments 2 and 3) is higher than that over the oxygen-containing seven-membered ring (Fragment 7) and its adjacent rings (Fragments 2 and 3). This disparity may give rise to differences in intermolecular stacking patterns and transfer integrals. Comprehensive analysis indicates that the introduction of nitrogen and boron atoms weakens their contribution to the HOMO, with boron atoms exerting a more pronounced effect.

**Reorganization energy**

Reorganization energy characterizes the energy changes during geometric structure relaxation of molecules upon electron gain or loss. As indicated by formula (1), reorganization energy is inversely proportional to carrier mobility—the smaller the reorganization energy, the more favorable for charge transfer. To this end, we employed the QM/MM model and calculated the hole reorganization energy and electron reorganization energy of the central molecule using the adiabatic potential energy surface method (APES) and normal mode analysis method (NMA) at the B3LYP/6-311G(2d,p) theoretical level, respectively. The results are presented in TABLE 5. As shown in TABLE 5, for **B1**, **B3**, and **B5**, the hole reorganization energy calculated by the NMA method is significantly higher than that obtained via the APES method. This discrepancy may arise from the mixing of low-frequency vibration modes during hole transport, which causes the harmonic oscillator approximation to fail and leads to overestimation of vibrational contributions. In contrast, electronic reorganization energies show substantial consistency between the two methods. The calculation results for other molecules exhibit good agreement across the two approaches, suggesting that their vibrations adhere to the harmonic oscillator model.

**TABLE 5** Hole reorganization energy ($\lambda_h$) and electron reorganization energy ($\lambda_e$) of the studied molecules in solid phase (unit: meV).

| Molecules | $\lambda_h$(AP) | $\lambda_h$(NM) | $\lambda_e$(AP) | $\lambda_e$(NM) |
|---|---|---|---|---|
| **A1** | 135 | 134 | 170 | 186 |
| **A2** | 157 | 156 | 184 | 198 |
| **A3** | 153 | 152 | 158 | 166 |
| **A4** | 153 | 153 | 200 | 201 |
| **B1** | 213 | 279 | 156 | 157 |
| **B2** | 191 | 215 | 159 | 160 |
| **B3** | 203 | 259 | 160 | 159 |
| **B4** | 164 | 169 | 159 | 160 |
| **B5** | 163 | 387 | 219 | 217 |
| **B6** | 162 | 163 | 234 | 236 |

Further analysis of the calculation results reveals that the hole reorganization energy of B-series molecules is generally higher than that of A-series molecules, with the magnitude of increase varying among specific molecules: **B1** and **B2** exhibit increases of 78 meV and 56 meV compared to **A1**, respectively; **B3** and **B4** show increases of 46 meV and 7 meV relative to **A2**, respectively; and **B5** and **B6** demonstrate increases of 10 meV and 9 meV compared to **A4**, respectively. This trend indicates that introducing sulfur or oxygen atoms between benzene rings to form seven-membered rings significantly elevates hole reorganization energy, with sulfur atoms exerting a more pronounced effect. The impacts of nitrogen and boron doping on hole reorganization energy differ between A-series and B-series molecules: **A2** (nitrogen-doped) shows a 22 meV increase in hole reorganization energy compared to **A1**, whereas **B3** and **B4** (nitrogen-doped) exhibit decreases of 10 meV and 27 meV relative to **B1** and **B2**, respectively. Similarly, **A4** (boron-doped) shows an 18 meV increase compared to **A1**, while **B5** and **B6** (boron-doped) exhibit decreases of 50 meV and 29 meV compared to **B1** and **B2**, respectively. Notably, although introducing nitrogen atoms (**A2**) or sulfur/oxygen atoms (**B1/B2**) individually in **A1** increases hole reorganization energy, simultaneous introduction of both nitrogen and sulfur/oxygen atoms (**B3/B4**) actually reduces reorganization energy, suggesting that nitrogen atoms can counteract the hole reorganization energy-increasing effect of sulfur/oxygen atoms. This non-resonant synergistic effect maintains the structural relaxation energy barrier during hole transport, keeping it within the standard parameter range for high-mobility materials (below 200 meV). Similarly, while introducing boron atoms alone (**A4**) increases reorganization energy, the simultaneous introduction of boron and sulfur/oxygen atoms (**B5/B6**) significantly suppresses the resonant effect of structural relaxation, reducing their reorganization energies to 163 meV and 162 meV, respectively.

To further investigate the origins of hole reorganization energy, the hole reorganization energy of each molecule was decomposed into its constituent vibrational modes, and the contribution of vibrational frequency $\omega_i$ to the hole reorganization energy $\lambda_h$ was analyzed, as depicted in FIGURE 7 and FIGURE 8.

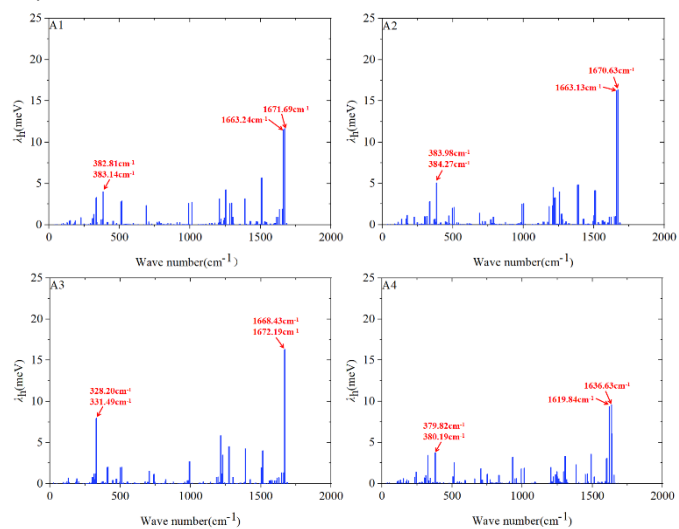

FIGURE 7. Contribution of vibrational modes to the hole reorganization energy ($\lambda_h$) of A-series molecules.

For analytical convenience, molecular vibration frequencies are categorized into two characteristic regions: 0–1000 cm$^{-1}$ (low frequency) and 1000–2000 cm$^{-1}$ (high frequency). Detailed analysis shows that **B1**, **B3**, and **B5** exhibit mixed vibrational modes in the low-frequency range, causing the harmonic oscillator approximation method to overestimate vibrational contributions. Consequently, in subsequent studies on the hole reorganization energy of **B1**, **B3**, and **B5**, excessively contributing low-frequency vibrational modes will be excluded to ensure calculation accuracy.

The calculation data reveal that the hole reorganization energy of A-series molecules is predominantly influenced by high-frequency vibrational modes. In the high-frequency range (1000–2000 cm$^{-1}$), in-plane stretching/shearing vibrations of carbon-carbon and carbon-nitrogen bonds, along with in-plane swinging vibrations of carbon-hydrogen bonds, make the most significant contributions. Specifically, **A2** (1663.13 cm$^{-1}$/1670.63 cm$^{-1}$) exhibits higher contributions than **A1** (1663.24 cm$^{-1}$/1671.69 cm$^{-1}$) (33 meV vs. 23 meV). Additionally, low-frequency range ( < 1000 cm$^{-1}$) overall stretching vibrations contribute more to **A2** (10 meV) than to **A1** (8 meV), leading to **A2** having a higher hole reorganization energy than **A1**.

Upon introducing nitrogen atoms into **A1**, the in-plane oscillation and stretching vibrations of carbon-nitrogen bonds become the primary contributing sources. Compared to **A1**, **A3** exhibits significantly increased contributions from high-frequency vibrations (1668.43 cm$^{-1}$/1672.19 cm$^{-1}$, 33 meV) and low-frequency overall stretching vibrations (328.20 cm$^{-1}$/331.49 cm$^{-1}$, 16 meV), with negligible contribution from the n-octyl branch. For boron-doped **A4**, in-plane stretching and swinging vibrations of carbon-carbon/boron-nitrogen bonds in the high-frequency region (1619.84 cm$^{-1}$/1636.63 cm$^{-1}$) contribute 19 meV, while low-frequency overall stretching vibrations (379.82 cm$^{-1}$/380.19 cm$^{-1}$) contribute 7 meV—both still exceeding those of **A1**.

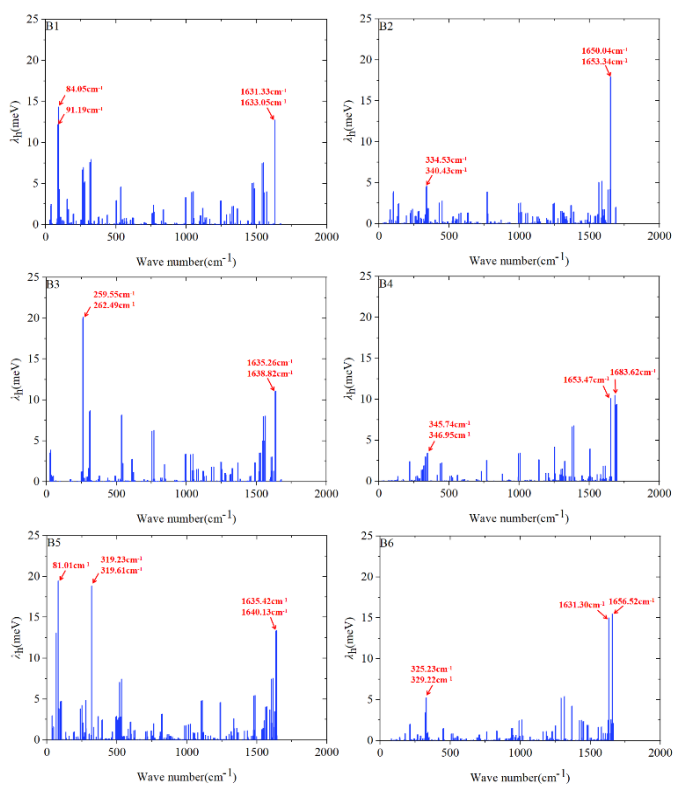

FIGURE 8. Contribution of vibrational modes to the hole reorganization energy ($\lambda_h$) of B-series molecules.

The hole reorganization energy of B-series molecules is predominantly influenced by the introduction of sulfur, oxygen, nitrogen, and boron atoms. Specifically, sulfur and oxygen atoms contribute primarily through out-of-plane oscillations in the low-frequency region and shear vibrations of carbon-sulfur or carbon-oxygen bonds in the high-frequency region. Nitrogen atoms mainly exhibit in-plane vibrations of carbon-nitrogen bonds, whereas boron atoms exert their effect via in-plane stretching vibrations that form boron-nitrogen bonds with nitrogen atoms. The key molecular vibrational characteristics are as follows.

**B1** and **B2** (parent molecules): In the high-frequency range (1600–1700 cm$^{-1}$), shear and in-plane stretching vibrations of carbon-sulfur/carbon-oxygen bonds and carbon-carbon/carbon-nitrogen bonds make the largest contributions, accounting for 25 meV (11.82%) and 36 meV (16.59%), respectively. The low-frequency overall stretching vibration contribution (9 meV) in **B2** is smaller than that in **B1**, leading to lower overall reorganization energy. **B3** and **B4** (nitrogen-doped molecules): Compared with **B1** and **B2**, their high-frequency contributions decrease to 22 meV and 21 meV, respectively, while low-frequency vibrational contributions are significantly suppressed (**B3**: 28 meV; **B4**: 38 meV), resulting in reduced reorganization energy. **B5** and **B6** (boron-doped molecules): relative to **B1** and **B2**, vibrational contributions further decrease, with high-frequency contributions of 13 meV and 30 meV, and low-frequency contributions of 16 meV and 39 meV, respectively. The reorganization energy follows a similar trend.

A comprehensive comparison between A-series and B-series molecules reveals that the introduction of sulfur and oxygen atoms introduces additional carbon-sulfur and carbon-oxygen bond vibrations, which largely account for the higher hole reorganization energy in B-series molecules. The formation of seven-membered rings by incorporating sulfur or oxygen atoms between benzene rings significantly elevates the hole reorganization energy of **B1** and **B2** due to these new vibrational modes. Notably, introducing nitrogen atoms (to form **B3** and **B4**) or boron atoms (to form **B5** and **B6**) into the edge benzene rings of **B1** and **B2** effectively mitigates this effect. Analysis shows that while the vibrations of carbon-nitrogen and boron-nitrogen bonds contribute modestly to reorganization energy, their presence significantly dampens the vibrational intensity of carbon-sulfur and carbon-oxygen bonds, thereby reducing the overall hole reorganization energy. This finding provides critical theoretical guidance for regulating reorganization energy in molecular design.

**Molecular stacking, transfer integration, and intermolecular interactions**

**Molecular stacking and transfer Integral**

In an ideal crystal, the charge transfer properties of materials are predominantly governed by the spatial arrangement and orientation of molecules in the solid state, with transfer integrals exhibiting significant sensitivity to the relative positions between molecules. This study selected **A1**, **A3**, **B1**, and **B2**—pyrrole-ring-containing molecules with known crystal structures—as the research objects, and conducted a detailed analysis of their molecular stacking patterns and spatial configurations of nearest-neighbor pairs (as shown in FIGURE 9 and FIGURE 10).

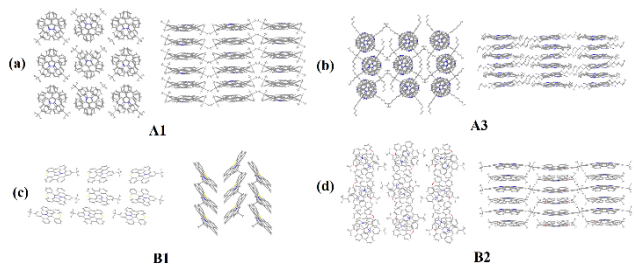

FIGURE 9. Packing motifs of **A1**, **A3**, **B1**, and **B2** in crystal structures.

The research results show that, in contrast to linear dibenzene, these polycyclic aromatic hydrocarbon molecules predominantly exhibit one-dimensional columnar stacking characteristics. Crystal structure analysis reveals that the molecular stacking mode is closely correlated with the depth of the bowl-shaped structure. **A1** and **A3** feature deeper bowl-shaped structures, **B1** exhibits a quasi-planar structure, and **B2** displays a shallower bowl-shaped structure.

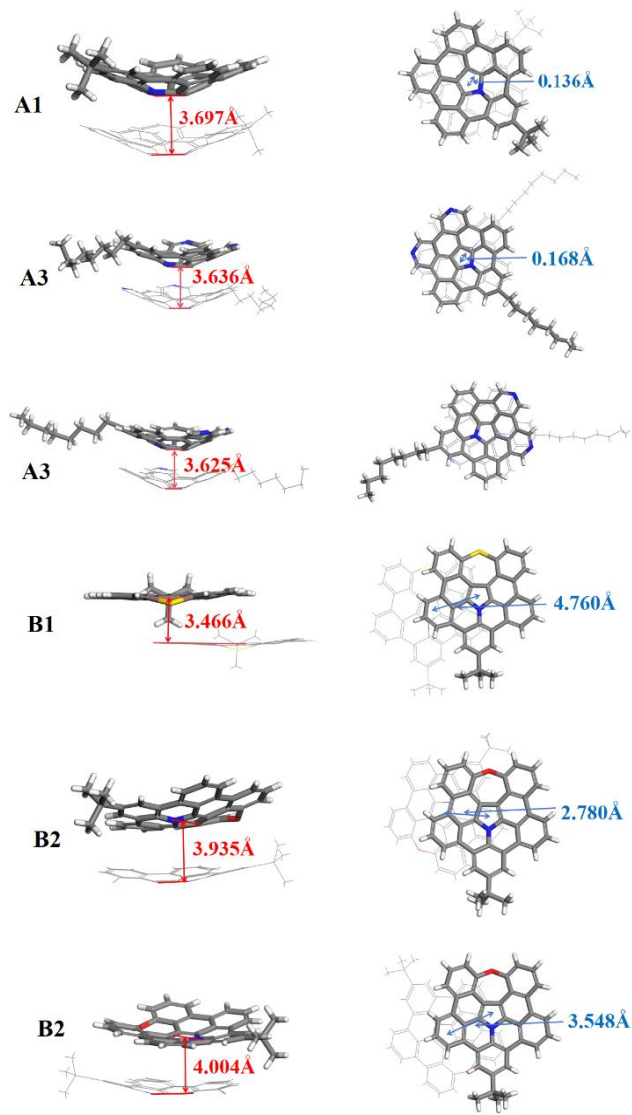

FIGURE 10. The nearest neighbor pairs of **A1**, **A3**, **B1**, and **B2** in the crystal structure (**A2**, **A4** have the same stacking pattern as **A1**, **B3**, **B5** have the same stacking pattern as **B1**, and **B4**, **B6** have the same stacking pattern as **B2**).

**A1** exhibits unidirectional concave-convex π-π columnar stacking with an intermolecular tilt angle of 2.5°, slip of 0.136 Å, and spacing of 3.697 Å. Although **A3** also forms unidirectional concave-convex π-π stacking, it displays two distinct stacking modes within the column: one with a tilt angle of 6.2° (negligible slip, spacing 3.625 Å) and another with a tilt angle of 4.4° (slip 0.168 Å, spacing 3.636 Å). **B1** features slightly slipped inclined face-to-face π-π stacking (slip 4.760 Å, spacing 3.466 Å), enabling significant orbital overlap. **B2** presents two slip stacking modes, both with a tilt angle of 14.9° but differing in slip distance: 3.548 Å (spacing 4.004 Å) and 2.780 Å (spacing 3.935

Å). In the crystal structure, similar to **A1**, both **A2** and **A4** exhibit unidirectional concave-convex π-π columnar stacking along the axis, with adjacent columns stacked in antiparallel orientations. In the column, the tilt angle between two adjacent molecules is 2.5°, the slip distance is 0.135 Å, and the intermolecular spacing is 3.697 Å. Similar to **B1**, **B3** and **B5** exhibit slightly slipped tilted face-to-face π-π stacking, with a neighboring intermolecular spacing of 3.466 Å, enabling significant orbital overlap and intermolecular face-to-face interactions. Analogous to **B2**, **B4** and **B6** display sliding π-π stacking structures. Two distinct stacking modes are observed in adjacent molecular pairs of **B2**: one pair features a 14.9° tilt angle, 3.548 Å slip distance, and 4.004 Å intermolecular spacing; the other pair also has a 14.9° tilt angle but with a 2.780 Å slip distance and 3.935 Å spacing. Overall, A-series molecules with deeper bowl-shaped structures tend to form compact unidirectional concave-convex π-π stacking, whereas B-series molecules with shallower bowl-shaped structures exhibit more pronounced slip-based π-π stacking patterns.

Carrier mobility is primarily governed by transfer integrals, which are determined by the relative spatial arrangements of adjacent molecules. Given that all molecules in this study exhibit P-type transport characteristics, the analysis focused on hole transfer integrals. Transfer integral values and corresponding transport pathways calculated using the PW91/TZP method are listed in TABLES S2–S5 of the supplementary information. The hole transfer integral distributions for selected charge transport pathways in **A1**, **A3**, **B1**, and **B2** are shown in FIGURE 11, FIGURE 12, FIGURE 13, and FIGURE 14. Since the sign of the transfer integral does not affect transfer rate analysis, only absolute values are discussed herein.

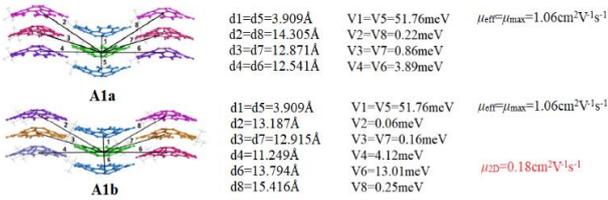

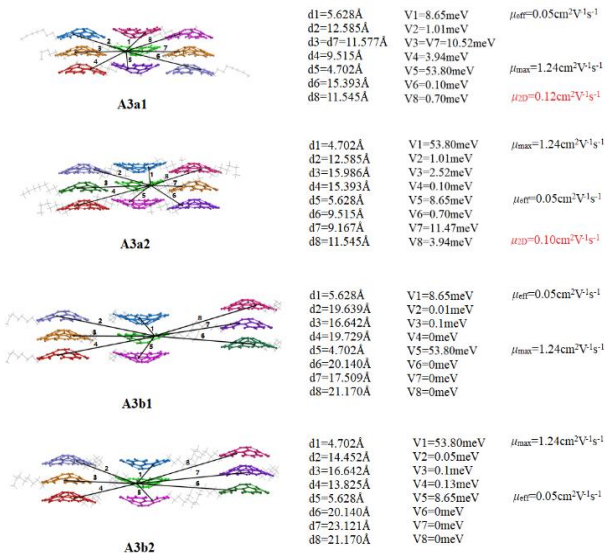

FIGURE 11. The centroid distance d of the main charge transfer path of **A1**, and the hole transfer integral V (absolute value), maximum one-dimensional hole mobility $\mu_{max}$, the most effective one-dimensional hole mobility $\mu_{eff}$ and the average two-dimensional hole mobility $\mu_{2D}$.

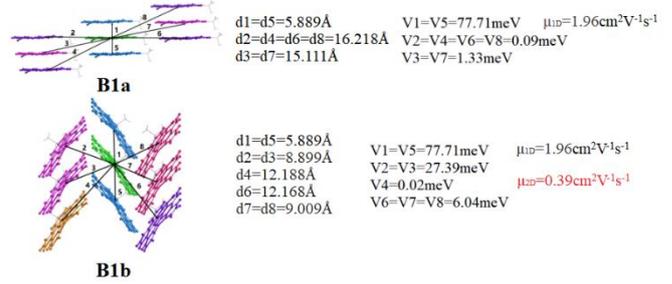

FIGURE 12. The centroid distance d of the main charge transfer path of **A3**, and the hole transfer integral V (absolute value), maximum one-dimensional hole mobility $\mu_{max}$, the most effective one-dimensional hole mobility $\mu_{eff}$ and the average two-dimensional hole mobility $\mu_{2D}$.

FIGURE 13. The centroid distance d of the main charge transfer path of **B1**, and the hole transfer integral V (absolute value), maximum one-dimensional hole mobility $\mu_{max}$, the most effective one-dimensional hole mobility $\mu_{eff}$ and the average two-dimensional hole mobility $\mu_{2D}$.

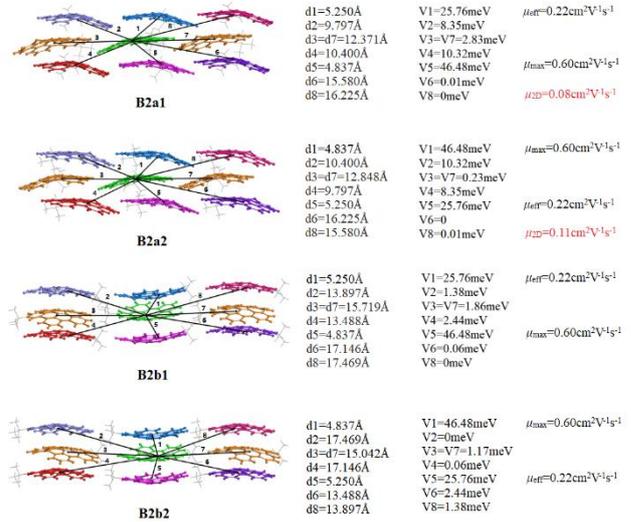

FIGURE 14. The centroid distance d of the main charge transfer path of **B2**, and the hole transfer integral V (absolute value), maximum one-dimensional hole mobility $\mu_{max}$, the most effective one-dimensional hole mobility $\mu_{eff}$ and the average two-dimensional hole mobility $\mu_{2D}$.

Molecular charge transport occurs across two dimensions (a and b): intra-layer and inter-layer. For **A3** and **B2**, the presence of two stacking modes within the same one-dimensional column enables further subdivision into a1, a2, b1, and b2 dimensions. The results indicate that the charge transfer characteristics of each molecule are predominantly governed by the centroid distance and the degree of molecular orbital overlap.

**A1** exhibits the highest hole transfer integral (51.76 meV) at a centroid distance of 3.909 Å, attributed to its minimal slip degree. For **A3**, the maximum hole transfer integral (53.80 meV) occurs in a molecular pair with a centroid distance of 4.702 Å. However, due to its asymmetric hole transport pathway, the most effective charge transport pathway corresponds to the molecular pair with a centroid distance of 5.628 Å, where the hole transfer integral is 8.65 meV. Notably, **B1** still exhibits the highest hole transfer integral (77.71 meV) at a centroid distance of 5.889 Å, primarily attributed to its exceptional intermolecular orbital overlap. Analysis of **B2** further corroborates this trend, with the maximum hole transfer integral (46.48 meV) observed in the molecular pair at a centroid distance of 4.837 Å. However, similar to **A3**, its asymmetric charge transfer pathway results in the most effective charge transport pathway corresponding to the molecular pair with a centroid distance of 5.250 Å, where the hole transfer integral is 25.76 meV. In this study, hole transfer integrals were calculated for both the charge transfer pathway of the nearest molecular pair and the most effective charge transfer pathway for each substituted derivative. The hole transfer integrals $V_{h(max)}$ and $V_{h(eff)}$ for the nearest molecular pair (i.e., the pair with the largest transfer integral) and the molecular pair corresponding to the most effective hole transfer pathway are listed in TABLE 6. The corresponding maximum HOMO overlap integrals and most effective HOMO overlap integrals are depicted in FIGURE 15.

**TABLE 6** The centroid distance (unit: Å), face-to-face distance (unit: Å), and maximum hole transfer integral $V_{h(max)}$ (unit: meV) of the nearest neighbor molecule pairs studied. The centroid distance (unit: Å), face-to-face distance (unit: Å), and most effective hole transfer integral $V_{h(eff)}$ (unit: meV) of the molecule pairs corresponding to the most effective transport path studied.

| Molecules | Centroid distance(Å) | Face-to-face distance(Å) | $V_{h(max)}$ | $V_{h(eff)}$ |
|---|---|---|---|---|
| A1 | 3.909 | 3.697 | 51.76 | 51.76 |
| A2 | 3.902 | 3.697 | 50.06 | 50.06 |
| A3 | 4.702 | 3.636 | 53.80 | - |
| A3 | 5.628 | 3.625 | - | 8.65 |
| A4 | 3.911 | 3.697 | 54.88 | 54.88 |
| B1 | 5.889 | 3.466 | 77.71 | 77.71 |
| B2 | 4.837 | 3.935 | 46.48 | - |
| B2 | 5.250 | 4.004 | - | 25.76 |
| B3 | 5.889 | 3.466 | 49.88 | 49.88 |
| B4 | 4.831 | 3.935 | 29.05 | - |
| B4 | 5.245 | 4.004 | - | 5.46 |
| B5 | 5.889 | 3.466 | 80.42 | 80.42 |
| B6 | 4.840 | 3.935 | 26.87 | - |
| B6 | 5.248 | 4.004 | - | 12.72 |

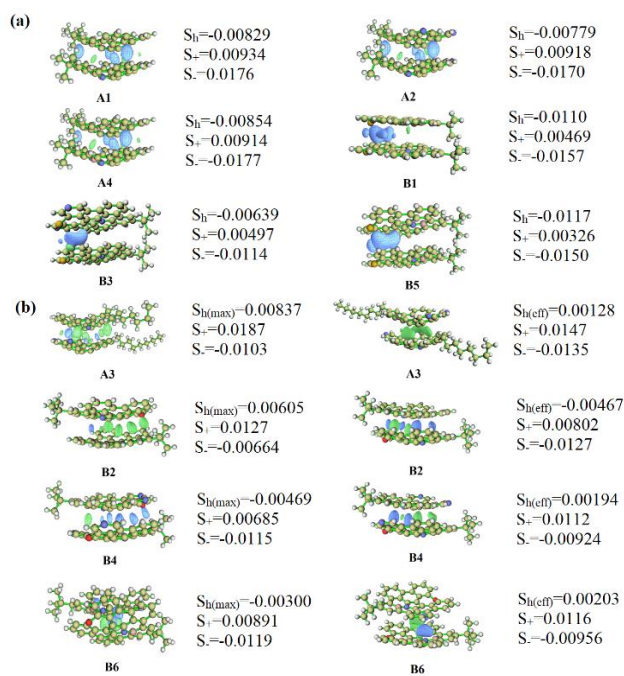

FIGURE 15. The HOMO overlap integral $S_{h(max)}$ of the maximum charge transfer channel of the molecule, as well as the HOMO overlap integral $S_{h(eff)}$ of the most effective charge transfer channel. $S_+$ represents the orbital overlap integral of the same phase, and $S_-$ represents the orbital overlap integral of the opposite phase (unit: a.u.).

The hole transfer integrals of **B3**, **B4**, and **B6** were reduced by 20.1%, 41.7%, and 53.2%, respectively, compared to the parent molecules **B1** and **B2**. Analysis of the HOMO distribution reveals that introducing nitrogen atoms significantly localizes the HOMO near sulfur or oxygen atoms, leading to decreases in both the maximum and most effective HOMO overlap integrals. This weakens the electron coupling effect and reduces both the maximum and most effective hole transfer integrals.

Notably, despite the introduction of nitrogen atoms in the comparison between **B1** and **B5**, **B5** maintains excellent hole transport performance (80.42 meV) due to its similar HOMO overlap integral distribution. Additionally, compared to **A4**, **B5** features smaller molecular spacing and larger maximum and most effective HOMO overlap integrals, resulting in a 46.5% increase in hole transfer integral.

**Intermolecular Interactions**

In order to investigate the rotation and slip mechanisms of molecules during the stacking process, weak interactions between molecules were calculated.

Firstly, the electrostatic potential diagram of a single molecule was calculated (FIGURE S1 in the supplementary information). By analyzing the distribution map of electrostatic potential (ESP), the nucleophilic sites (positive potential points) and electrophilic sites (negative potential points) of molecules can be identified, which are crucial for intermolecular interactions.[62]

For A-series molecules, the peripheral benzene ring regions of **A1** and **A4** exhibit significant electronegativity characteristics. In contrast, **A2** and **A3** introduce nitrogen atoms, but these nitrogen atoms exhibit different electrical distributions in the ESP spectrum: the nitrogen atoms on the pyrrole ring show significant positive charge, while the introduced nitrogen atoms exhibit strong electronegativity. The difference in electrical distribution is closely related to the redistribution of molecular structure and its electron cloud.

For B-series molecules, ESP analysis reveals that the sulfur and oxygen atom regions in the seven membered ring structure exhibit strong electronegativity. It is worth noting that **B3** and **B4** not only exhibit significant electronegativity on the introduced nitrogen atom, but also exhibit strong electronegativity on the sulfur and oxygen atoms on their seven membered rings.

In addition, all nitrogen atoms on the pyrrole ring in the B-series molecules exhibit a positive charge, which is related to the unique curved structure of the molecules, the curved configuration causes the electron cloud to redistribute on the pyrrole ring, resulting in the central nitrogen atom showing a positive charge in the ESP spectrum. The intermolecular interaction energy is a quantitative indicator of the strength of intermolecular interactions, and its magnitude can be used to measure the relative stability of dimers. The formation of molecular stacking can also be analyzed by analyzing the interaction between the dimers.

TABLE 7 lists the interaction energy ($E_{SCS-SAPT0}$) and energy decomposition components of important transport channel molecules in A-series and B-series molecules, including exchange repulsion energy ($E_{exch}$), electrostatic energy ($E_{elec}$), induction energy ($E_{ind}$), and dispersion energy ($E_{disp}$).

Among them, exchange repulsion energy can play a repulsive role; electrostatic energy, induction energy, and dispersion energy play an attractive role. The ratio of attractive energy to repulsive energy can be used to measure the stability of molecular dimer. Generally, the closer the ratio is to 2.0, the more stable the molecular dimer is. From the calculation results, it can be found that dispersion energy dominates the interaction energy between A-series molecules and B-series molecules (**A1**: 51.48%; **A2**: 52.21%; **A3**: 50.21%; **A4**: 49.33%; **B1**: 48.42%; **B2**: 51.89%; **B3**: 51.47%; **B4**: 52.49%; **B5**: 48.35%; **B6**: 51.45%). The energy changes of molecules after substitution were analyzed and compared. There is a certain comparison between absolute value and relative value.

**TABLE 7** Energy decomposition components (unit: kcal/mol) and proportion of nearest neighbor molecular pair and molecular pair corresponding to the most effective transport path in the studied molecules.

| Molecules-cetroid distance(Å) | $E_{SCS-SAPT0}$ | $E_{elec}$ | $E_{ind}$ | $E_{disp}$ | $E_{exch}$ | $E_{att}/E_{rep}$ |
|---|---|---|---|---|---|---|
| **A1**-3.909 | -54.25 | -19.85 / 12.52% | -4.94 / 3.12% | -81.62 / 51.48% | 52.15 / 32.89% | 2.04 |
| **A2**-3.902 | -52.73 | -18.53 / 12.68% | -4.60 / 3.15% | -76.33 / 52.21% | 46.73 / 31.97% | 2.13 |
| **A3**-4.702 | -55.35 | -22.17 / 13.56% | -5.16 / 3.16% | -82.09 / 50.21% | 54.08 / 33.07% | 2.02 |
| **A3**-5.628 | -57.06 | -23.59 / 14.34% | -5.25 / 3.19% | -81.93 / 49.81% | 57.31 / 32.66% | 2.06 |
| **A4**-3.911 | -53.07 | -21.00 / 12.95% | -6.61 / 4.08% | -79.99 / 49.33% | 54.53 / 33.63% | 1.97 |
| **B1**-5.889 | -37.63 | -16.98 / 14.19% | -3.81 / 3.18% | -57.95 / 48.42% | 40.95 / 34.21% | 1.92 |
| **B2**-4.837 | -44.37 | -16.10 / 13.00% | -3.74 / 3.02% | -64.25 / 51.89% | 39.72 / 32.08% | 2.12 |
| **B2**-5.250 | -45.16 | -17.69 / 14.17% | -3.72 / 2.98% | -63.59 / 50.94% | 39.83 / 31.91% | 2.13 |
| **B3**-5.889 | -34.84 | -12.16 / 11.25% | -3.67 / 3.40% | -55.62 / 51.47% | 36.61 / 33.88% | 1.95 |
| **B4**-4.831 | -43.99 | -15.64 / 13.80% | -3.52 / 3.11% | -59.52 / 52.49% | 34.70 / 30.60% | 2.27 |
| **B4**-5.245 | -44.35 | -16.90 / 14.86% | -3.51 / 3.09% | -58.62 / 51.55% | 34.68 / 30.50% | 2.28 |
| **B5**-5.889 | -34.42 | -14.90 / 12.48% | -4.28 / 3.58% | -57.74 / 48.35% | 42.51 / 35.59% | 1.81 |
| **B6**-4.840 | -43.64 | -15.74 / 13.20% | -4.35 / 3.65% | -61.35 / 51.45% | 37.80 / 31.70% | 2.15 |
| **B6**-5.248 | -42.17 | -16.78 / 14.02% | -4.03 / 3.37% | -60.10 / 50.23% | 38.74 / 32.38% | 2.09 |

Compared with B-series molecules, A-series molecules have a larger total interaction energy, indicating that A-series nearest neighbor molecules are more stable. In the B-series molecules, the interaction energies of oxygen-containing **B2**, **B4**, and **B6** are all greater than those of sulfur-containing **B1**, **B3**, and **B5**, indicating that the nearest neighbor pairs of oxygen-containing B-series molecules are more stable compared to sulfur-containing B-series molecules.

Compared to **A1**, the total interaction energy and various sub energies of **A2** with nitrogen atoms introduced are decreased ($E_{SCS-SAPT0}$: from 54.25 kcal/mol to 52.73 kcal/mol; $E_{elec}$: from 19.85 kcal/mol to 18.53 kcal/mol; $E_{ind}$: from 4.94 kcal/mol to 4.60 kcal/mol; $E_{disp}$: from 81.62 kcal/mol to 76.33 kcal/mol; $E_{exch}$: from 52.15 kcal/mol to 46.73 kcal/mol), indicating that the introduction of nitrogen atoms reduces steric hindrance and reduces spatial overlap, the total interaction energy and various sub energies of **A3** with nitrogen atoms and long branched chains increased ($E_{SCS-SAPT0}$: from 54.25 kcal/mol to 55.35 kcal/mol; $E_{elec}$: from 19.85 kcal/mol to 22.17 kcal/mol; $E_{ind}$: from 4.94 kcal/mol to 5.16 kcal/mol; $E_{disp}$: from 81.62 kcal/mol to 82.09 kcal/mol; $E_{exch}$: from 52.15 kcal/mol to 54.08 kcal/mol), indicating that octyl increases steric hindrance and spatial overlap. Compared with **B1**, the total interaction energy and various sub energies of **B3** with nitrogen atoms introduced are decreased ($E_{SCS-SAPT0}$: from 37.63 kcal/mol to 34.84 kcal/mol; $E_{elec}$: from 16.98 kcal/mol to 12.16 kcal/mol; $E_{ind}$: from 3.81 kcal/mol to 3.67 kcal/mol; $E_{disp}$: from 57.95 kcal/mol to 55.62 kcal/mol; $E_{exch}$: from 40.95 kcal/mol to 36.61 kcal/mol), indicating that the introduction of nitrogen atoms reduces steric hindrance and reduces spatial overlap; compared with **B2**, the total interaction energy and various sub energies of **B4** with nitrogen atoms introduced are reduced ($E_{SCS-SAPT0}$: from 44.37 kcal/mol to 43.99 kcal/mol; $E_{elec}$: from 16.10 kcal/mol to 15.64 kcal/mol; $E_{ind}$: from 3.74 kcal/mol to 3.52 kcal/mol; $E_{disp}$: from 64.25 kcal/mol to 59.52 kcal/mol; $E_{exch}$: from 39.72 kcal/mol to 34.70 kcal/mol), indicating that the introduction of nitrogen atoms reduces steric hindrance and spatial overlap. The above analysis indicates that introducing nitrogen atoms will reduce the steric hindrance and spatial overlap between molecular pairs.

Comparing **A1** with **A4** obtained by introducing boron atomn, the total interaction energy of **A4** is reduced, electrostatic energy, induction energy and exchange repulsion energy is increased; dispersion energy and is reduced ($E_{SCS-SAPT0}$: from 54.25 kcal/mol to 53.07 kcal/mol; $E_{elec}$: from 19.85 kcal/mol to 21.00 kcal/mol; $E_{ind}$: from 4.94 kcal/mol to 6.61 kcal/mol; $E_{disp}$: from 81.62 kcal/mol to 79.99 kcal/mol; $E_{exch}$: from 52.15 kcal/mol to 54.53 kcal/mol). Comparing **B1** with **B5** obtained by introducing boron atom, the total interaction energy of **B5** is reduced; electrostatic energy is reduced; induction energy is increased; dispersion energy is reduced; exchange repulsion energy is increased ($E_{SCS-SAPT0}$: from 37.63 kcal/mol to 34.42 kcal/mol; $E_{elec}$: from 16.98 kcal/mol to 14.90 kcal/mol; $E_{ind}$: from 3.81 kcal/mol to 4.28 kcal/mol; $E_{disp}$: from 57.95 kcal/mol to 57.74 kcal/mol; $E_{exch}$: from 40.95 kcal/mol to 42.51 kcal/mol). Comparing **B2** with **B6** obtained by introducing boron atoms, the total interaction energy of **B6** is reduced; electrostatic energy is reduced; induction energy is increased; dispersion enerey is reduced and exchange repulsion enerey is increased ($E_{SCS-SAPT0}$: from 44.37 kcal/mol to 43.64 kcal/mol; $E_{elec}$: from 16.10 kcal/mol to 15.74 kcal/mol; $E_{ind}$: from 3.74 kcal/mol to 4.35 kcal/mol; $E_{disp}$: from 64.25 kcal/mol to 61.35 kcal/mol; $E_{exch}$: from 39.72 kcal/mol to 37.80 kcal/mol). That indicates that boron atoms have different effects on the stacking structures of **A1**, **B1**, and **B2**. Due to the different molecular structures and dimer configurations of these molecules, the comparison between absolute values is limited. Relative values can sometimes provide valuable information.

Compared to **A1**, the contribution ratio of electrostatic energy, induction energy, and dispersion energy of **A2** with nitrogen atoms increases ($E_{elec}$: from 12.52% to 12.68%; $E_{ind}$: from 3.12% to 3.15%; $E_{disp}$: from 51.48% to 52.21%); the contribution of exchange repulsion energy decreases ($E_{exch}$: from 32.89% to 31.97%); the contribution ratio of electrostatic energy, induction energy, and exchange repulsion energy of **A3** with long branched chains and nitrogen atoms increases ($E_{elec}$: from 12.52% to 13.56%; $E_{ind}$: from 3.12% to 3.16%; $E_{exch}$: from 32.89% to 33.07%), the contribution of dispersion energy decreases ($E_{disp}$: from 51.48% to 50.21%). Compared to **A1**, the contribution ratio of electrostatic energy, induction energy, and exchange repulsion energy of **A4** with boron atom increases ($E_{elec}$: from 12.52% to 12.95%; $E_{ind}$: from 3.12% to 4.08%; $E_{exch}$: from 32.89% to 33.63%), the contribution of dispersion energy decreases ($E_{disp}$: from 51.48% to 49.33%). Compared with **B1**, the contribution ratio of the induction energy and dispersion energy of **B3** with nitrogen atoms increases ($E_{ind}$: from 3.18% to 3.40%; $E_{disp}$: from 48.42% to 51.47%); the contribution ratio of electrostatic energy and exchange repulsion energy decreases ($E_{elec}$: from 14.19% to 11.25%; $E_{exch}$: from 34.21% to 33.88%). Compared with **B2**, the proportion of electrostatic energy, induction energy, and dispersion energy of **B4** with nitrogen atoms increases ($E_{elec}$: from 13.00% to 13.80%; $E_{ind}$: from 3.02% to 3.11%; $E_{disp}$: from 51.89% to 52.49%); the contribution ratio of exchange repulsion energy decreases ($E_{exch}$: from 32.08% to 30.60%). Compared with **B1**, the contribution ratio of the induction energy and exchange repulsion energy of **B5** with boron atoms increases ($E_{ind}$: from 3.18% to 3.58%; $E_{exch}$: from 34.21% to 35.59%); the contribution ratio of electrostatic energy and dispersion energy decreases ($E_{elec}$: from 14.19% to 12.48%; $E_{disp}$: from 48.42% to 48.35%). Compared with **B2**, the contribution ratio of electrostatic energy and induction energy of **B6** with boron atoms increased ($E_{elec}$: from 13.00% to 13.20%; $E_{ind}$: from 3.02% to 3.65%); the contribution ratio of dispersion energy and exchange repulsion energy decreases ($E_{disp}$: from 51.89% to 51.45%; $E_{exch}$: from 32.08% to 31.70%).

Through energy decomposition analysis of intermolecular interaction energy, it was found that the introduction of nitrogen or boron atoms significantly affects the characteristics of intermolecular interactions. Specifically, **A2** and **A3** with nitrogen atoms exhibit different interaction modes: the proportion of attractive energy of **A2** increases (**A1**: 67.12%; **A2**: 68.04%), while the proportion of exchange repulsion energy decreases (**A1**: 32.89%; **A2**: 31.97%), indicating a smaller molecular stacking distance (**A1**: 3.909 Å; **A2**: 3.902 Å); on the other hand, **A3** exhibits the opposite trend of interaction, with a decrease in the proportion of attractive energy (**A1**: 67.12%; **A3**: 66.93%) and an increase in the proportion of repulsive energy (**A1**: 32.89%; **A3**: 33.07%).

The increase in substituent chain length may lead to an enhancement of its repulsive effect, which may result in larger intermolecular stacking distances (**A1**: 3.909 Å; **A3**: 4.702 Å, 5.628 Å), accompanied by significant molecular tilt and slip. In addition, compared with **A1**, **A4**, which introduces boron atom, have similar interaction characteristics, and their enhanced repulsion (**A1**: 32.89%; **A4**: 33.63%) may also lead to an increase in the molecular stacking distance (**A1**: 3.909 Å; **A4**: 3.911 Å).

The analysis of the effect of introducing heteroatoms on the interaction between benzene rings shows that the introduction of sulfur and oxygen atoms significantly changes the molecular stacking behavior. Compared with **A1**, the repulsive effect of sulfur-containing atoms on **B1** is enhanced (**A1**: 32.89%; **B1**: 34.21%), resulting in an increase in molecular stacking distance (**A1**: 3.909 Å; **B1**: 5.889 Å), an increase in slip distance (**A1**: 0.136 Å; **B1**: 4.760 Å), and a decrease in tilt angle. Although **B2** containing oxygen atoms exhibits stronger attraction (**A1**: 67.12%; **B2**: 67.91%), its molecular spacing shows a similar increasing trend as sulfur atoms (**A1**: 3.909 Å; **B2**: 4.837 Å), and the slip distance is relatively small (**B1**: 4.760 Å; **B2**: 2.780 Å, 3.548 Å). It is worth noting that although **B2** has a higher proportion of attractive energy than **B1** (**B1**: 65.79%; **B2**: 67.91%), its molecular spacing did not decrease as expected. This anomalous phenomenon may be due to the bowl shaped depth and relatively high electronegativity of oxygen-containing seven membered ring monomers, resulting in slight slip but large rotation of the two molecules, and the steric hindrance effect of the n-Bu group at the tail is greater, which is the result of the combined effect of the bowl shaped depth and electronegativity brought by the introduction of oxygen atoms and the steric hindrance of substituents.

The study on the regulation of molecular stacking behavior by the introduction of nitrogen and boron atoms shows that the type and relative position of heteroatoms significantly affect intermolecular interactions. Compared with **B1**, the proportion of attraction energy of **B3** containing nitrogen atoms increases (**B1**: 65.79%; **B3**: 66.12%), which may lead to a decrease in molecular spacing; similarly, **B4** showed the same trend compared to **B2** (**B2**: 67.91%; **B4**: 69.40%). However, the introduction of boron atoms presents different regulatory effects: **B5** exhibits stronger repulsion compared to **B1** (**B1**: 34.21%; **B5**: 35.59%), indicating larger molecular spacing; compared to **B2**, the enhanced attraction of **B6** (**B2**: 67.91%; **B6**: 68.30%) may promote a decrease in intermolecular distance. These results not only reveal the differential regulatory mechanisms of nitrogen and boron atoms on intermolecular interactions, but also provide new insights for precise regulation of molecular stacking behavior through heteroatom engineering.

This study employed the Hirshfeld surface analysis method to quantitatively characterize the crystal structures of **A1**, **A3**, **B1**, and **B2**. By visualizing parameters including $d_{norm}$, shape index (SI), and curvature, and integrating two-dimensional fingerprint spectra (as shown in FIGURE S2 and FIGURE S3 of the supplementary information), a comprehensive assessment of the crystal structural features was achieved.

Hirshfeld shape index spectra serve as a powerful tool for analyzing the characteristics of intermolecular contacts. Notably, the complementary red-blue triangular regions within these spectra typically signify pronounced π···π stacking interactions, which reflect the close-contact mode of molecular surfaces in these specific regions.

The types of intermolecular interactions—such as hydrogen bonds, halogen bonds, and π···π stacking—can be systematically identified through color-coded analysis of two-dimensional fingerprint plots. The visualization advantage of this method not only enhances the understanding of molecular interaction modes but also holds significant application value in crystal structure prediction and new material development.

Contact configuration analysis reveals that sulfur and oxygen atoms exert a notable influence on molecular stacking, with sulfur atoms demonstrating a more pronounced regulatory effect in the crystal structure owing to their larger atomic radius and stronger intermolecular interactions. Shape index analysis indicates that π···π and C–H···π interactions play a pivotal role in molecular stacking, where the π···π and C–H···π interactions of **A1**, **A3**, and **B2** are located in the center and peripheral regions of the molecule, respectively, forming typical columnar stacking, while the symmetrical distribution of these two types of interactions on both sides of the molecule causes **B1** to form a unique tilted π-stacking structure..

**Charge transport properties of molecules**

**Carrier mobility**

In accordance with the Brownian motion model, a molecule was selected from the crystal structure as the charge transfer starting point, with all neighboring molecules considered as potential next jump sites. A dynamic Monte Carlo method was employed to conduct a total of 2000 random jump experiments, where the occurrence time of each jump and the diffusion distance were recorded. Subsequently, through analysis of data from these 2000 experiments, the diffusion coefficient $D$ was determined via linear regression. The charge transfer rate of molecule pairs in the transport channel formed by each jump was calculated, and based on the previously calculated hole reorganization energy under solid-phase model conditions, the transfer rate of each transport path for the studied molecule was computed using the Marcus formula.

On this basis, the hole carrier mobility of each transport path for the studied molecule was calculated using Einstein's formula (see TABLEs S6–S9 in the supplementary information), with the maximum one-dimensional hole mobility and the most effective one-dimensional hole mobility presented in TABLE 8. For molecules whose crystal structures were predicted via simulated construction, the maximum one-dimensional hole mobility ($\mu_{h(max)}$) of the primary transport path and the hole mobility ($\mu_{h(eff)}$) of the most effective one-dimensional transport path were also calculated, as listed in TABLE 8.

**TABLE 8** The centroid distance (unit: Å), face to face distance (unit: Å) and maximum hole mobility $\mu_{h(max)}$ (unit: cm$^2$V$^{-1}$s$^{-1}$), the centroid distance (unit: Å), face to face distance (unit: Å) and the most effective hole mobility $\mu_{h(eff)}$ (unit: cm$^2$V$^{-1}$s$^{-1}$) of the most effective charge transfer path of the molecules under study.

| Molecules | Centroid distance(Å) | Face-to-face distance(Å) | $\mu_{h(max)}$ | $\mu_{h(eff)}$ |
|---|---|---|---|---|
| **A1** | 3.909 | 3.697 | 1.06 | 1.06 |
| **A2** | 3.902 | 3.697 | 0.67 | 0.67 |
| **A3** | 4.702 | 3.636 | 1.24 | - |
| **A3** | 5.628 | 3.625 | - | 0.05 |
| **A4** | 3.911 | 3.697 | 0.82 | 0.82 |
| **B1** | 5.889 | 3.466 | 1.96 | 1.96 |
| **B2** | 4.837 | 3.935 | 0.60 | - |
| **B2** | 5.250 | 4.004 | - | 0.22 |
| **B3** | 5.889 | 3.466 | 0.46 | 0.46 |
| **B4** | 4.831 | 3.935 | 0.16 | - |
| **B4** | 5.245 | 4.004 | - | 0.01 |
| **B5** | 5.889 | 3.466 | 3.49 | 3.49 |
| **B6** | 4.840 | 3.935 | 0.28 | - |
| **B6** | 5.248 | 4.004 | - | 0.08 |

Through theoretical calculations and analysis of hole mobility, we have revealed the mechanism by which different molecular structures affect charge transport performance.

Specifically, the molecular pairs in **A1** with a center-of-mass distance of 3.909 Å show minimal slippage and the optimal overlapping degree, yielding a hole mobility of 1.06 cm$^2$V$^{-1}$s$^{-1}$. In contrast, the most effective transport path of A3 features a centroid distance of 5.628 Å with negligible slippage, but a larger intermolecular tilt angle causes the hole mobility to decrease to 0.05 cm$^2$V$^{-1}$s$^{-1}$.

The centroid distance of the transport path of **B1** is 5.889 Å, and the intermolecular distance is only 3.466 Å. Combined with a large hole transfer integral, its hole mobility can reach up to 1.96 cm$^2$V$^{-1}$s$^{-1}$. Due to the large intermolecular distance (4.004 Å) of the transport path with a centroid distance of 5.250 Å and a small HOMO overlap integral, the hole mobility of **B2** decreases to 0.22 cm$^2$V$^{-1}$s$^{-1}$.

A comparison of **B1** and **B2** reveals that **B1** exhibits a symmetrical herringbone stacking structure, where the nearest-neighbor molecules show substantial HOMO overlap—an attribute conducive to charge transfer. In contrast, **B2** features a loose and highly slipped π-π stacking structure, characterized by significant rotation between nearest-neighbor molecular pairs and minimal HOMO overlap, both of which hinder hole transport. As a result, **B1** demonstrates higher hole mobility.

In summary, intermolecular distance, centroid distance, and slip degree are key factors influencing hole mobility. By consolidating the precomputed data for each studied molecule, a comprehensive analysis was performed on one-dimensional hole mobility and its influencing factors for the most effective molecular transport pathways, with results summarized in TABLE 9.

Analysis of the tabulated data provides an integrated assessment of molecular one-dimensional hole mobility and its influencing factors. Notably, the primary determinants of the most effective one-dimensional hole mobility differ between A-series and B-series molecules: hole reorganization energy exerts a more substantial influence on A-series molecules, whereas the hole transfer integral plays a dominant role in governing the mobility of B-series molecules.

**TABLE 9** The hole reorganization energy $\lambda_h$ (unit: meV), the most effective hole overlap integral $s_{h(eff)}$ (unit: a.u.), the most effective hole transfer integral $V_{h(eff)}$ (unit: meV), and the most effective one-dimensional hole mobility $\mu_{h(eff)}$ (unit: cm$^2$V$^{-1}$s$^{-1}$) of studied molecules.

| Molecules | $\lambda_h$ | $S_{h(eff)}$ | $V_{h(eff)}$ | $\mu_{h(eff)}$ |
|---|---|---|---|---|
| **A1** | 135 | -0.00829 | 51.76 | 1.06 |
| **A2** | 157 | -0.00779 | 50.06 | 0.67 |
| **A3** | 153 | 0.00128 | 8.65 | 0.05 |
| **A4** | 153 | -0.00854 | 54.88 | 0.82 |
| **B1** | 213 | -0.0110 | 77.71 | 1.96 |
| **B2** | 191 | -0.00467 | 25.76 | 0.22 |
| **B3** | 203 | -0.00639 | 49.88 | 0.46 |
| **B4** | 164 | 0.00194 | 5.46 | 0.01 |
| **B5** | 163 | -0.0117 | 80.42 | 3.49 |
| **B6** | 162 | 0.00203 | 12.72 | 0.08 |

Specifically, upon introducing two nitrogen atoms into **A1** to form **A2**, the HOMO overlap integral and hole transfer integral slightly decrease, while enhanced vibrations in both low-frequency and high-frequency regions lead to an increase in hole reorganization energy, ultimately causing a reduction in hole mobility. **A3** exhibits a similar phenomenon of vibration enhancement, but the increased tilt angle of nearest-neighbor molecules reduces the HOMO overlap integral and hole transfer integral, resulting in an unexpected increase in its most effective one-dimensional hole mobility. Notably, the introduction of boron atoms into **A4** enhances the HOMO overlap integral, leading to a larger hole transfer integral; however, the increased vibrational contribution subsequently reduces hole mobility. In the B-series, introducing nitrogen atoms into **B1** to form **B3** weakens vibrations and decreases hole reorganization energy, yet a reduction in the HOMO overlap integral results in diminished hole mobility. Conversely, introducing boron atoms to form **B5** increases the HOMO overlap integral while weakening vibrational contributions, both elevating the hole transfer integral and reducing hole reorganization energy—ultimately causing a significant enhancement in hole mobility. For **B2** derivatives, both **B4** and **B6** show reduced vibration, leading to a decrease in hole reorganization energy. However, the decrease in the HOMO overlap integral of nearest-neighbor pairs causes a reduction in the hole transfer integral, ultimately resulting in a decline in the most effective one-dimensional hole mobility. Systematic studies on sulfur-containing seven-membered ring molecules have demonstrated that modifications in molecular structure exert a significant regulatory effect on their hole mobility. Specifically, compared with **B1**, the introduction of nitrogen atoms in **B3** leads to a decrease in one-dimensional hole mobility, while the comparison between **B1** and **B5** reveals that the introduction of boron-nitrogen units effectively enhances one-dimensional hole mobility. Further investigation shows that compared with **A4**, **B5** achieves a more pronounced improvement in one-dimensional hole mobility by introducing boron-nitrogen units and sulfur atoms between benzene rings.

Compared with B-series molecules, the introduction of sulfur and oxygen atoms into A-series molecules induces changes in their geometric and electronic structures, thereby altering intermolecular interactions and stacking modes, which ultimately leads to variations in transfer integrals and migration rates. Among these, introducing sulfur and oxygen atoms—particularly sulfur atoms—exerts a more profound influence on molecular structure and stacking modes than introducing nitrogen or boron atoms. Moreover, the combined introduction of sulfur and boron atoms exhibits an even greater impact. This discovery provides a critical theoretical foundation for optimizing charge transport performance through precise regulation of molecular structures, offering important theoretical guidance for enhancing charge transport efficiency in molecular design.

It is observed that doping nitrogen or boron atoms into the ring of **A1** leads to the effect of reorganization energy surpassing that of the transfer integral: the increase in reorganization energy causes a decrease in one-dimensional hole mobility. For the series of molecules, in **B1** and **B5** with sulfur atoms introduced between rings, the contribution of the transfer integral outweighs that of the reorganization energy—although the reorganization energy increases, the rise in the transfer integral drives an increase in one-dimensional hole mobility. In contrast, for **B2**, **B4**, and **B6** with oxygen atoms introduced between rings, the decrease in the transfer integral and the increase in reorganization energy collectively result in a reduction in one-dimensional hole mobility.

## Conclusions

This study addresses the stability limitations of π-conjugated P-type semiconductors by investigating charge transport regulation in nonplanar bowl-shaped polycyclic aromatic hydrocarbons (PAHs) using density functional theory and molecular dynamics simulations. Through atomic doping and structural modification of prototype molecules (**A1**, **A2**, **A3**, **B1**, and **B2**), a series of derivatives (**A4**, **B3**, **B4**, **B5**, and **B6**) were designed, yielding key insights.

(1) Bowl-shaped depth control.

Deeper bowl-shaped PAHs form tightly packed concave-convex columnar π-stacks with minimal slip, enhancing hole mobility (governed by transfer integrals) compared to shallower, looser structures.

(2) Heteroatom doping.

Introduction of S/O atoms widens the bandgap and enhances optical stability.

In sulfur/oxygen-bridged seven-membered rings, N/B doping suppresses hole reorganization energy ( < 200 meV), meeting high-mobility criteria. Orbital redistribution modifies π-stacking: **B1** exhibits tilted stacks, while **B5** (with boron-nitrogen units and sulfur-bridging) achieves superior columnar stacking and hole mobility (3.49 vs. 1.96 $cm^2 \cdot V^{-1} \cdot s^{-1}$ in **B1**).

This work establishes a structure–property relationship linking heteroatom doping, molecular geometry, and stacking configuration, proposing strategies based on bowl depth and heteroatom incorporation to design stable, high-mobility organic semiconductors. Future research should focus on experimental validation of these predictions and explore assembly dynamics to optimize device performance.

## Author contributions

Heng-yu Jin: Writing – review & editing, Writing – original draft, Software, Resources, Project administration, Methodology, Investigation, Formal analysis, Data curation, Conceptualization. Xiao-qi Sun: Writing – original draft, Formal analysis, Data curation. Gui-ya Qin: Writing – review & editing, Visualization, Validation. Zhi-peng Tong: Methodology, Conceptualization. Rui Wang: Supervision, Software, Resources. Qi Zhao: Investigation. Ai-Min Ren: Writing – review & editing, Writing – original draft, Resources, Investigation, Funding acquisition. Jing-Fu Guo: Software, Resources, Methodology, Funding acquisition.

## Conflicts of interest

There are no conflicts to declare.

## Data availability

The calculations in this work based on Density Functional Theory (DFT) were performed by Gaussian09 (version D.01) program package. Gaussian: https://gaussian.com/. The charge transfer integrals were performed using Amsterdam Modeling Suite (abbreviation AMS, formerly ADF). AMS: https://fermitech.com.cn/. Multiwfn code: http://sobereva.com/multiwfn/. The remaining data supporting the findings of this study are available from the corresponding author upon request.

## Acknowledgements

This work was supported by the Key Research and Development Project of Jilin Provincial Department of Science and Technology (No. 20240302015GX), Natural Science Foundation of Jilin Province of China (No. 20240101167JC) and the Natural Science Foundation of China (Nos. 21473071, 21173099 and 20973078). This work is also supported by HPC of BUCT.

## Notes and references